\pdfoutput=1

\IfFileExists{latexml.sty}{\RequirePackage{latexml}}{\newif\iflatexml\latexmlfalse}

\iflatexml 
\else 
\fi 

\documentclass[conference]{IEEEtran}  \newcommand \equalizelastpgcols  {0mm}

\usepackage[utf8]{inputenc}
\usepackage[T1]{fontenc}

\def\equalizetitlepagecols {0mm} 
\IEEEoverridecommandlockouts

\usepackage{amsmath,amsfonts} \usepackage{array}
\iflatexml 
\else 
\usepackage{tgcursor}
\usepackage{algorithm} 
\fi 

\usepackage{comment}
\usepackage{alltt} 
\usepackage{upquote} 

\usepackage{graphicx} 

\usepackage{amsthm} 

\newlength{\bufferhline} 

\def\myblue{blue}
\def\myblack{black}

\iflatexml 
\usepackage{color}
\usepackage{soul}
\setulcolor{blue}
\usepackage[colorlinks=true]{hyperref}
\usepackage[square,numbers]{natbib}
\newcommand{\shortdoi}[1]{\href{https://doi.org/#1}{\ul{\color{\myblack}\texttt{doi:#1}}}}
\newcommand{\wvcite}[1]{\citep[{\color{\myblue}{$\Rightarrow$}}][]{#1}}
\newcommand{\wvref}[1]{\hyperref[#1]{{\color{\myblue}{$\Rightarrow$}}{\color{\myblack}{\ref{#1}}}}}
\newcommand{\wveqref}[1]{\hyperref[#1]{{\color{\myblue}{$\Rightarrow$}}{\color{\myblack}{\eqref{#1}}}}}
\def\mybegineqstar#1{\begin{equation}\label{#1}}
\def\myendeqstar{\end{equation}}
\def\mytag#1{}
\else 
\usepackage[normalem]{ulem}
\usepackage[svgnames]{xcolor}
\usepackage[colorlinks=true,allcolors=\myblack]{hyperref}
\usepackage[table]{hypcap}
\newcommand{\shortdoi}[1]{{\color{\myblue}\uline{\color{\myblack}\href{https://doi.org/#1}{\texttt{doi:#1}}}}}
\newcommand{\wvcite}[1]{{\color{\myblue}\uline{{\color{\myblack}\cite{#1}}}}}
\newcommand{\wvref}[1]{{\color{\myblue}\uuline{{\color{\myblack}\ref{#1}}}}}
\newcommand{\wveqref}[1]{{\color{\myblue}\uline{{\color{\myblack}\eqref{#1}}}}}
\def\mybegineqstar#1{\hbox{\refstepcounter{equation}\label{#1}}\begin{equation*} }
\def\myendeqstar{\end{equation*}}
\def\mytag#1{\tag{#1}}
\fi 

\newcommand \doichoice [1] {\shortdoi{#1}} 

\iflatexml 
\else 
\usepackage{afterpage}
\usepackage{footnotehyper}
\fi 

\newcommand \nbdd {\nobreakdash}
\newcommand{\Z}{\hphantom{0}}
\newlength{\myrowskip}

\iflatexml 
\newcommand{\myTwoColCaptionvspace} [1][]{}
\newcommand{\myTwoColSectionvspace} [1][]{} 
\newcommand{\myTwoColTableLiftvspace} [1][]{}
\newcommand{\mydepthbox}{\vrule height0pt depth10pt width0pt}

\newcommand \myOtimes {\odot}
\def\mybox#1{$#1_{9}$}
\def\mytabcolsepA{2.0em}
\else 
\newcommand{\myTwoColCaptionvspace} [1][-2.0mm]{\vspace{#1}}
\newcommand{\myTwoColSectionvspace} [1][-2.0mm]{\vspace{#1}} 
\newcommand{\myTwoColTableLiftvspace} [1][-3.0mm]{\vspace{#1}} 
\newcommand{\mydepthbox}{\vrule height0pt depth0pt width0pt}

\def\mybox#1{\boxed{#1}}
\newcommand \myOtimes {\hat{\otimes}}
\def\mytabcolsepA{0.15em}
\fi 

\iflatexml 

\newtheorem{theorem}{Theorem}
\newtheoremstyle{mystyle}%
  {}
  {}
  {\noindent\itshape}
  {}
  {\bfseries\large}
  {~:}
  {\newline}
  {\noindent\thmname{#1}\thmnumber{ #2}\thmnote{ (#3)}} 
\theoremstyle{mystyle}
\newtheorem{theorem}{Theorem}
\newtheorem{algorithm}{Algorithm(s)}
\def\myTheorem{theorem}
\def\myabstract{myabstractenv}
\def\mykeywords{mykeywordsenv}
\def\myproof{myproofenv}
\else 

\def\myTheorem{theorem}
\def\myabstract{abstract}
\def\mykeywords{IEEEkeywords}
\def\myproof{proof}
\fi 

\def  \BeginAlgFigure {\begin{algorithm}}
\def  \EndAlgFigure {\end{algorithm}}
\def \AlgFigure {Algorithm}

\usepackage[english]{babel} 

\makeatletter
\let\ORIbbl@fixname\bbl@fixname
\def\bbl@fixname#1{%
  \@ifundefined{languagealias@\expandafter\string#1}
    {\ORIbbl@fixname#1}
    {\edef\languagename{\@nameuse{languagealias@#1}}}%
}
\newcommand{\definelanguagealias}[2]{%
  \@namedef{languagealias@#1}{#2}%
}
\let\BIBforeignlanguage\foreignlanguage
\makeatother

\definelanguagealias{en}{english}

\begin{document}
\iflatexml 
\def\mythankstext{\copyright The author, dunning@bgsu.edu, %
grants copyright permissionsas specified in\\
Creative Commons License CC BY-SA 4.0. (See Endnote 3)\\ 
\\
}
\else 
\def\mythankstext{\copyright The author  grants copyright permissions as specified in %
Creative Commons License CC BY-SA 4.0.  Attributions should reference this manuscript. %
This version typeset with current program code as of \today.
}
\fi 
\newcommand \thankstext {\mythankstext}

\title{\LARGE Length 3 Check Digit Codes with Grouped Tags \\[-0.2ex]
and Disjoint Coding Applications\\[-1.2ex]}

\author{\IEEEauthorblockN{Larry A. Dunning%
\iflatexml 
\\[0.5ex]%
\else 
Prof. Emeritus}%
\fi 
\thanks{\thankstext}
\IEEEauthorblockA{\small Dept. of Computer Science\\
Bowling Green State University\\
Bowling Green, Ohio 43403 \\
Email: dunning@bgsu.edu} \\[-3.0ex]} 

\makeatletter
\def\myDoublePARstart{\@IEEESAVECMDIEEEPARstart}
\makeatother

\maketitle

\iflatexml 
\else 
\makeatletter
\let\lastpagenumber8 
\def\ps@plain{%
  \let\@mkboth\@gobbletwo
   \renewcommand{\@oddhead}{}%
  \renewcommand{\@evenhead}{}%
  \renewcommand{\@evenfoot}{\reset@font\small\hfil\thepage{}~of~\lastpagenumber \hfil}%
  \renewcommand{\@oddfoot}{\@evenfoot}%
}
\pagestyle{plain} 
\makeatother
\thispagestyle{plain}
\fi 

\begin{\myabstract}
In 1969 J. Verhoeff provided the first examples of a decimal error detecting code 
using a single check digit to provide protection  against all single, 
transposition and adjacent twin errors.
The three codes he presented are length 3\nbdd-digit codes with 2 information digits.  
To date, the existence of a length 4\nbdd-digit code 
with 3 information digits and these properties remains an open question.  
Existence of a 4\nbdd-digit code would imply the existence of 10 disjoint 3\nbdd-digit codes. 
Apparently, no pair of such disjoint 3\nbdd-digit codes is known.  
Phonetic errors, where 2 digit pairs of the forms X0 and 1X are interchanged, 
are language dependent, but can often be eliminated.  
Alternate 3\nbdd-digit codes are developed here 
which enhance the level of protection beyond Verhoeff's codes 
while still optionally providing protection against phonetic errors. 
Through almost-disjoint coding schemes, 
it is shown how copies of these new codes 
can fill in the gap between such 3 and 4\nbdd-digit codes. 
The results are extended to other useful alphabet sizes such as 26 and 36 
with stronger permutation of digits error detection, 
and to ``tag codes" where digits are grouped.
\end{\myabstract}

\begin{\mykeywords}
Decimal and alphanumeric error detection, transposition errors, 
twin errors, phonetic errors, cyclic errors.
\end{\mykeywords}

\enlargethispage{\equalizetitlepagecols}

\myTwoColSectionvspace
\section{Verhoeff's Curious 3-digit Decimal Code}
\iflatexml 
In
\else 
\myDoublePARstart{I}{n} 
\fi 
his 1969 monograph Jacobus Verhoeff \wvcite{Verhoeff1969}
presented 
three variations of the  ``curious 3\nbdd-digit decimal code", derived from a block design.
shown in Table~\wvref{VerhoeffRegular}.
Doing this in the most direct manor 
yielded the code given in Table~\wvref{VerhoeffRegular}.
Each table entry gives the middle digit $m$ of the codeword throughout this paper, 
allowing a simpler correspondence between properties of the table 
and the requirements for detecting each error type.  
A potentially confused pair of codewords will be counted as an error.
An error in just one of the three digits 
is termed a \textit{single error} or \textit{transcription error}. 
Note that the single error detecting property implies that 
any two digits determine the remaining digit.  
This property, as do the other detection properties, 
also implies a corresponding  \textit{erasure correction} capability.  
Detection of single errors requires 
that each row and each column of the table be a permutation.  
Detection of \textit{twin errors}, 
where two identical adjacent digits change identically, 
requires that these permutations each have at most one fixed point.  
Detection of \textit{transposition errors}, 
where two differing adjacent digits become interchanged, 
requires that these permutations have no cycles of length~2. 
\iflatexml 
\else 
\def\noteone{In the original tables on pages 40-41 of \wvcite{Verhoeff1969}, 
the table ``resulting" from the block design 
is erroneously interchanged with the ``interchanged" table (not shown) in which the codeword ``\texttt{109}" appeared as ``\texttt{100}".}
\fi 
\begin{table}[ht] \label{VerhoeffRegular}
\iflatexml 
\caption[]{Verhoeff's Block Design Code$\textsuperscript{\wvref{foot1}}\mydepthbox$\\}
\else 
\caption[]{Verhoeff's Block Design Code$\color{blue}\uuline{\protect\footnote{\noteone}}$}
\def\noteone{In the original tables on pages 40-41 of \wvcite{Verhoeff1969}, 
the table ``resulting" from the block design 
is erroneously interchanged with the ``interchanged" table (not shown) in which the codeword ``\texttt{109}" appeared as ``\texttt{100}".}
\fi 
\centering%
\myTwoColCaptionvspace
\setlength{\tabcolsep}{0.5em}
\begin{tabular}{c|cccccccccc}
$M$&0&1&2&3&4&5&6&7&8&9\\
\hline
\rule{0pt}{2ex}   
0&0&3&1&2&9&4&5&6&7&8\\
1&2&1&3&0&5&8&7&4&9&6\\
2&3&0&2&1&7&6&9&8&5&4\\
3&1&2&0&3&8&9&4&5&6&7\\
4&5&7&9&6&4&1&8&2&3&0\\
5&6&4&8&7&0&5&2&9&1&3\\
6&7&9&5&8&3&0&6&1&4&2\\
7&8&6&4&9&1&3&0&7&2&5\\
8&9&5&7&4&6&2&3&0&8&1\\
9&4&8&6&5&2&7&1&3&0&9\\
\end{tabular}
\end{table}
\begin{table}[ht]\label{VerhoeffIrregular}
\myTwoColTableLiftvspace
\iflatexml 
\caption{Verhoeff's Irregular Code\textsuperscript{\wvref{foot1}}\mydepthbox}
\else 
\caption{Verhoeff's Irregular Code} %
\fi 
\centering
\myTwoColCaptionvspace
\setlength{\tabcolsep}{0.5em}
\begin{tabular}{c|cccccccccc}
$M$&0&1&2&3&4&5&6&7&8&9\\
\hline
\rule{0pt}{2ex}   
0&0&3&4&9&6&7&5&8&2&1\\
1&5&1&0&2&8&3&9&6&7&4\\
2&7&6&2&4&1&0&8&9&3&5\\
3&1&5&8&3&7&6&4&0&9&2\\
4&2&9&7&5&4&8&1&3&0&6\\
5&6&7&9&0&3&5&2&4&1&8\\
6&3&8&1&7&5&9&6&2&4&0\\
7&9&4&5&8&2&1&0&7&6&3\\
8&4&0&6&1&9&2&3&5&8&7\\
9&8&2&3&6&0&4&7&1&5&9\\
\end{tabular}
\myTwoColCaptionvspace
\end{table}
\iflatexml 
\else 
\afterpage{\setcounter{footnote}{1}\footnotetext{In the original tables on pages 40-41 of \wvcite{Verhoeff1969}, 
the table ``resulting" from the block design 
is erroneously interchanged with the ``interchanged" table (not shown) in which the codeword ``\texttt{109}" appeared as ``\texttt{100}".}}
\fi 
Table \wvref{VerhoeffRegular}  contains 
the codewords \texttt{302} and \texttt{132} yielding a \textit{right phonetic error} 
as well the codewords \texttt{230} and \texttt{213} yielding a \textit{left phonetic error}.  
Rather than displaying an interchanged table that removes the phonetic errors therein, 
a better irregular code from \wvcite{Verhoeff1969} appears in Table~\wvref{VerhoeffIrregular}.
All three codes detect single, transposition errors and twin errors.  
They also detect all \textit{jump twin errors}, 
where a codeword \texttt{aba} is interchanged with \texttt{cbc}, 
requiring that the main diagonal be a permutation.   
Detection of \textit{jump transposition errors}, 
where \texttt{abc} is exchanged with \texttt{cab}, 
is provided by making the table completely asymmetric about the main diagonal. 
The irregular code also detects all phonetic errors. 
Verhoeff \wvcite{Verhoeff1969} notes that 
none of the codes protects against \textit{triple errors} 
where \texttt{aaa} is interchanged with \texttt{bbb} 
and hence they are ``perhaps not recommendable"  in practice.  
The main cited advantage of the irregular code is that 
it provides protection against 82.8\% of the \textit{cyclic errors}, 
where \texttt{abc} and \texttt{bca} are interchanged, 
while the block design code and interchanged code provide protection against none. 
If all cyclic errors were detected, 
then the codes would be \textit{permutation error} detecting and 
knowledge of the multiset of digits in a codeword 
is then sufficient to specify the codeword. 
Table~ \wvref{ErrorTypes} lists the error types and conditions for their detection.
\begin{table}[tb]
\caption{Detection Conditions for Table M 3-digit Error Patterns\mydepthbox} \label{ErrorTypes}
\centering
\myTwoColCaptionvspace
{
\setlength{\tabcolsep}{\mytabcolsepA}
\begin{tabular}{@{}l|*{4}{c|}}
&Error Types
&All Rows
&All Columns
&Main Diagonal \\[\myrowskip]
\hline
\rule{0pt}{2.2ex}   
&Single(*)&$\neg \; abc\leftrightarrow abd$&$\neg \; acd\leftrightarrow bcd$& \\
&&permutation&permutation&  \\[1mm]
\hline 
\rule{0pt}{2.2ex} 
&Transposition&$\neg \; abc\leftrightarrow acb$&$\neg \: abc\leftrightarrow bac$& \\
&&No 2-cycles&No 2-cycles&   \\[1mm]
\hline
\rule{0pt}{2.2ex} 
&Twin&$\neg \; abb\leftrightarrow acc$&$\neg \;aac\leftrightarrow bbc$& \\
& &$\leq 1$ fixed point&$\leq 1$ fixed point&  \\[1mm]
\hline
\rule{0pt}{2.2ex} 
&J. Transposition&  &  & $\neg \; abc\leftrightarrow cab$\\
 &  &  &  &asymmetric \\[1mm]
\hline
\rule{0pt}{2.2ex} 
&J. Twin& &  & $\neg \; aca\leftrightarrow bcb$ \\
&&  &  & permutation \\[1mm]
\hline
\rule{0pt}{2.2ex}  
&Triple&  &  &$\neg \; aaa\leftrightarrow bbb$ \\
&&  & &$ \leq 1$ fixed point \\[1mm]
\hline
\rule{0pt}{2.2ex}  
&Phonetic Left&\multicolumn{3}{c|}{$\forall e: 0\neq M(M(1,e),e) 
\implies \forall x \: \neg\;1xe\leftrightarrow x0e$}\\[1mm]
\hline
\rule{0pt}{2.2ex} 
&Phonetic Right&\multicolumn{3}{c|}{$\forall b: 1\neq M(b,M(b,0)) 
\implies \forall x \: \neg \;b1x\leftrightarrow bx0$}\\[1mm]
\hline
\rule{0pt}{2.2ex} 
&Cyclic&\multicolumn{3}{c|}{ \( \forall a,b: b \neq M((Ma,b),a)
\implies \forall x \: \neg \; axb \leftrightarrow xba \) } \\[1mm]
\hline
\rule{0pt}{2.2ex} 
&\multicolumn{4}{c|}{(*) $Function: M(b,e) 
\implies \neg \; abd\leftrightarrow acd$}\\[1mm]
\hline
\end{tabular}}\end{table}
\myTwoColSectionvspace[-6.0mm]
\section{An Improved 3-digit Decimal Code}
The code shown in Table \wvref{Code47374800} 
provides all the protections cited for the previous codes.  
The code shown protects against all phonetic errors.  
In addition, all triple errors are detected 
as the main diagonal permutation has only one fixed point.  
There are 9 cyclic errors that are not detected 
compared to 16 undetected cyclic errors in the irregular code 
of \wvcite{Verhoeff1969} shown in Table \wvref{VerhoeffIrregular}.  
Because both codes handle all transposition error types, 
the possible permutation errors are restricted to the cyclic errors.
Section~\wvref{OtherBases} will address some points with more formality, 
while implementation issues are addressed first. 

\begin{table}[ht]
\myTwoColTableLiftvspace
\caption{A Decimal Code Using $GF(3^2)$\mydepthbox}\label{Code47374800}
\centering
\myTwoColCaptionvspace
\setlength{\tabcolsep}{0.5em}
\begin{tabular}{c|cccccccccc}
$M$&0&1&2&3&4&5&6&7&8&9\\
\hline
\rule{0pt}{2ex}   
0&3&8&1&2&4&9&7&0&5&6\\
1&2&4&6&9&0&5&3&8&1&7\\
2&7&0&5&3&9&1&2&4&6&8\\
3&1&3&8&6&2&4&5&7&9&0\\
4&6&2&4&5&7&0&9&3&8&1\\
5&5&7&0&1&3&8&6&9&4&2\\
6&8&1&9&4&6&2&0&5&7&3\\
7&9&6&2&0&5&7&8&1&3&4\\
8&0&9&7&8&1&3&4&6&2&5\\
9&4&5&3&7&8&6&1&2&0&9\\
\end{tabular}
\end{table}

\myTwoColSectionvspace
\section{The Finite Field Computations}
A correspondence between the decimal digits `0' to `8' and the elements of $GF(3^2)$ 
will be used to create a base 9 code with the desired properties.  
The 9 by 9 table given by that code will be augmented 
by an additional row and an additional column.  
The digit `9' will then be substituted into the table yielding the desired code.  
If a chosen correspondence retains the field integers `0', `1' and `2', 
the results  herein will hold.
 
\def\myU{\_} 
 While the codes and their computation are completely described elsewhere in this paper,
\AlgFigure~\wvref{alg:html} can greatly facilitate the use of the codes.
When displayed, in a PDF reader or by a browser e.g. Adobe\textsuperscript{\textregistered}  or Chromium\textsuperscript{\textregistered} based,
the code can be exported by copying from the screen image
and pasted into an html text file.
The functions given in \AlgFigure~\wvref{alg:html}\nbdd-(A--B) 
represent the correspondence chosen for use in this paper.  
As the field's multiplicative group is cyclic, the field is simply specified by giving 
the powers and logarithms of the nonzero members 
with respect to a primitive root $4$ as arrays $pwA$ and $lgA$.  
Finite field operators $\oplus$, $\otimes$ and $\ominus$
are computed by the functions 
\(ad( \myU, \myU)\), \(ml(\myU, \myU)\) and \(sb(\myU,\myU)\).
See Section~\wvref{construction} to print their operation tables if desired.
Note that $I(\_)$ calculates the greatest integer function. 
The these operators or functions are also used later for $GF(5^2)$ calculations, 
except that the arrays $pwA$ and $lgA$ are then calculated using the function $mlP(\_)$ 
which multiplies a field element by a primitive element in $GF(5^2)$. 
The ECMAScript\textsuperscript{\textregistered} 
or JavaScript\textsuperscript{\textregistered} functions 
for computing code table values of \AlgFigure~\wvref{alg:html}\nbdd-(A\nbdd-D) 
have been translated into other programming languages for verifications. 
There, array operations such as \textit{map} and \textit{findIndex} may require coding loops.
Do note the use of the ternary operator 
\mbox{``\textit{boolean} ? \textit{value-if-true} : \textit{value-if-false"}}.
The table display and I/O code, \AlgFigure~\wvref{alg:html}\nbdd-(E), 
would need complete replacement in another environment.

The functions are encapsulated in minimal html above and below.  
They have been tested with several web browsers 
and allow easy generation of the codes developed herein. 
Code readability has been sacrificed to  
\iflatexml 
ensure fit into a single column in the PDF version of this manuscript.
\else 
allow it to fit into a single column. 
\fi 
The scripting language allows code to be evaluated at runtime, e.g. with $eval(\_)$.
Thus, enter  \mbox{\texttt{let x=4; ad(ad(ml(x,x),x),2)}}, into the web page input box and
press the ``=" button to view `0' in the display box 
confirming that $4\in GF(3^2)$ is a primitive element 
with minimal polynomial $x^{2}\oplus x\oplus 2$~\wvcite{Maurer2013}. 
\iflatexml 
\noindent\hrule
\begin{algorithm}[Algorithms for 10, 26, and 36 Letter Codes\textsuperscript{\wvref{foot2}\label{alg:html}}]
\end{algorithm}
{\par %
\ttfamily
\newcommand\bslash{\symbol{`\\}}
\newcommand\gr{\textasciigrave}
\newcommand\sltr[1]{\textquotesingle#1\textquotesingle}
\vspace{\bufferhline}  
\begin{alltt}
<html><head><meta charset="UTF-8">
<style>#OuE\{font-family: monospace;\}</style>
<title>Length 3 Check Digit Codes-ES6</title>
<script> help=()=>hp; Q=10;/*4,9,25,26,36,37*/ 
var urlstr=document.URL+\gr{}"?Q=$\{Q\}"\gr{};
  Q=+urlstr.match(/Q=([0-9]+)/)[1]; S=\sltr{ }; 
var _ =undefined, id=x=>x, I=x=>x|0; nl="\bslash{}n"; 
ta=n=>n<36?(Q%25<2?n+10:n).toString(36):\sltr{-}; 
AN=0; PK="P"; ts=x=>AN?x:ta(x); 
seq=n=>[...Array(n).keys()]; /*[0,..,n]*/
hp="Add to URL ?Q=N for 4,9,25,26,36,37\bslash{}n"; 
/*(A)*/ var q=9,G=8,Z=3,K=3,P=7,B=4,E=7;
var pwA = [ 1, 4, 6, 7, 2, 8, 3, 5 ];  
var lgA = [-1, 0, 4, 6, 1, 7, 2, 3, 5]; 
if(Q%25<2)\{G=24;Z=5; K=1;P=1;B=11;E=18; q=25; 
  mlP=x=>5*((I(x/5)+x%5)%5)+(3*I(x/5))%5; 
  pwA=seq(G).map(F=i=>(i==0)?1:mlP(F(i-1))); 
  lgA=seq(q).map(x=>pwA.findIndex(v=>v==x));\}; 
/*(B)*/ hp+="ad(x,y), ml,sb avail Q=4,9,25\bslash{}n"; 
  ad=(u,v,z=Z)=>(u+v)%z+z*((I(u/z)+I(v/z))%z); 
  ml=(u,v)=>(u*v==0)?0:pwA[(lgA[u]+lgA[v])%G]; 
  sb=(u,v)=>ad(u,ml((Z-1),v)); 
Mq=(b,e)=>sb(K,ad(ml(B,b),ml(E,e))); 
R=()=>sb(K,ml(B,P)); C=()=>ad(K,ml(E,P)); 
rwcl=(x,F)=>(x==q)?q:ad(x,F()); 
M=(b,e)=>(b==q)?rwcl(e,R):((e==q) ?  
  rwcl(b,C) : ((P==sb(b,e))?q:Mq(b,e)) ); 
/*(C)*/ ml4=x=>[0,2,3,1][x];  
M4=(b,e)=>ad(P,ml4(ad(b%4,ml4(e%4),2)),2); 
if(Q==36)\{p7=x=>((x%18==7)?x+18:x)%36; 
  M36=(b,e)=>(9*M4(b,e)+64*Mq(b%9,e%9))%Q; 
  M=(b,e)=>p7(M36(p7(b),p7(e))) \}; 
if(Q==4)\{M=M4;hp+="Fixed in GF(4): K,B,E\bslash{}n"; 
ml=(x,y)=>[0,1,2,3,3,_,1,_,_,2][x*y];q=4;Z=2\}; 
/*(D)*/ if(Q==37)\{ Q=36; K=1; B=10; E=26; 
M37=(b,e)=>((K-B*b-E*e)%37+37)%37; 
  M=M37; hp+="GF(37) K4,B,E are Fixed\bslash{}n";\}; 
if(Q==9||Q==25)M=Mq; if(Q<5||Q>35)K=(P=1);  
if(Q>26||Q==4)PK="K4";if(Q==25||Q==9)PK="_";
/*(E)*/ set=(k=K,p=P,b=B,e=E)=>(K=k,P=p, 
  B=b,E=e, \gr{}[K,$\{PK\},B,E]=[$\{[K,P,B,E]\}]\gr{}); 
Te=(b,m)=>(S=\sltr{e},(sQ.find(e=>M(b,e)==m))??36); 
sQ=seq(Q);SQ=sQ.map(ts);tm=s=>[s,...SQ]; 
tbl=T=>sQ.map(b=>(sQ.map(e=>ts(T(b,e))))); 
csv=t=>t.map(e=>e.join(",")).join(nl); 
lbl=t=>tm(S).map((x,i)=>[x,...[SQ,...t][i]]); 
show=(f=id,T=M,s=\sltr{m})=>(S=s,csv(f(tbl(T)))); 
hp+=\gr{}Size $\{Q\}: set() ... set(K,$\{PK\},B,E)\bslash{}n\gr{}
+ "m: See Code show() labeled: show(lbl)\bslash{}n"
+ "e: show(_,Te)/show(lbl,Te) AN=1 -> #s\bslash{}n"
+ "For Q=4,9,25 e.g. show(lbl,ml,\sltr{*})\bslash{}n"
+ "MJS()-More JS in myMJS.js";ul="myMJS.js"; 
MJS=()=>(js=document.createElement(\sltr{script}), 
  js.src=ul, document.body.append(js), ul); 
V=()=>document.getElementById("OuE").value= 
  eval(document.getElementById("InE").value);
document.write(\gr{}Current Alphabet Size $\{Q\}\gr{});
</script></head><body>See 2023 arXiv paper
<A HREF="https://doi.org/10.48550/arXiv.2312.12116">DOI</A>
<h4>Alpha. Size 10,26,36 Code Calculator</h4> 
<br><input type="text"id="InE"size=99 
  placeholder="Enter: help(), press = >"/><br> 
<input type="button"value="="onclick="V()"/> 
<br><textarea id="OuE" readonly  
  rows=37, cols=99/></textarea></body></html>
\end{alltt}
\par
} 
\noindent\hrule
\else 
\fi 
\myTwoColSectionvspace
\section{Construction of the Decimal Codes}\label{construction}
For simplicity codewords will be represented as $b \, m \, e$ 
designating their \textit{b}eginning, \textit{m}iddle (table entry), and \textit{e}nd digits.  
The $GF(q)$ code, $q=3^2$ for the present, from which the decimal code will be constructed, 
will have codewords satisfying 
\mybegineqstar{eq:M9}
\begin{split}
&B\otimes b \:\; \oplus \:\; 1 \otimes m \:\; \oplus \:\; E\otimes e \; = \; K \\
&M_q(b,e)\stackrel{\text{def}}{=} K \ominus (B \otimes b\; \oplus\; E \otimes e) 
\end{split}\mytag{$\theequation$}
\myendeqstar
where $B$, $E$ and $K$ are chosen with restrictions from $GF(q)$:%
\iflatexml 
\begin{equation}\label{eq:BEK}
\begin{align}
&0 \notin \{ B, E \} & \implies & \text{~~No single errors} \\
&0 \notin \{ B\oplus 1, E\oplus1, B\oplus E \} & \implies & \text{~~No (jump) twin}  \\
&0 \notin \{ B\ominus 1, E\ominus 1, B\ominus E \} & \implies & \text{~~No (jump) transposition} \\
&0 =  B\oplus 1\oplus E  \: \wedge \: K \neq 0 & \implies &\text{~~No triple words} 
\end{align}
\end{equation}
\else 
\hbox{\refstepcounter{equation}\label{eq:BEK}}%
\begin{align}
&0 \notin \{ B, E \} & \implies & \text{~~No single errors} \nonumber \\
&0 \notin \{ B\oplus 1, E\oplus1, B\oplus E \} & \implies & \text{~~No (jump) twin}\tag{\theequation}  \\
&0 \notin \{ B\ominus 1, E\ominus 1, B\ominus E \} & \implies & \text{~~No (jump) transposition}%
\nonumber \\
&0 =  B\oplus 1\oplus E  \: \wedge \: K \neq 0 & \implies &\text{~~No triple words} \nonumber
\end{align}
\fi 
The last of these restrictions is required because 
the codeword \texttt{999} will be added to the code in the conversion to decimal.  
The constant $K$ can be freely chosen from the nonzero field elements. 
Computer search limited  choices for $B$ and $E$ to:
\begin{align*} 
(B, E) \: \in \: \{ \;(3, 8), (4, 7), (5, 6), (6, 5), (7, 4), (8, 3)\; \}  
\end{align*}
The Frobenius automorphism $F(x) = x^3$ reduces the viable choices 
to the first three in the list, which are equivalent to the last three in order. 
The parameters $B=4, \; E=7, \; K=3$ were used in the construction 
of the code in Table \wvref{Code47374800},
which is derived from the $GF(3^2)$ code with these parameters 
as shown in Table \wvref{Code4737}.  
A code with similar properties over the group $Z_3 \times Z_3$ 
is given as Example 21 by Mullen and Shcherbacov \wvcite{mullen2004}.
\begin{table}[htb]
\myTwoColTableLiftvspace
\caption{Extending to a Decimal Code Using $GF(3^2)$\mydepthbox} \label{Code4737}
\centering
\myTwoColCaptionvspace
{
\setlength{\tabcolsep}{0.15em}
\begin{tabular}{lc|ccccccccc|c}
&$M$&$e=0$&$e=1$&$e=2$&$e=3$&$e=4$&$e=5$&$e=6$&$e=7$&$e=8$&$e=9$\\
\hline
\rule{0pt}{3ex}   
&$b=0$&3&8&1&2&4&\mybox{6}&7&0&5&$c_0$=6\\
&$b=1$&2&4&6&\mybox{7}&0&5&3&8&1&$c_1$=7\\
&$b=2$&7&0&5&3&\mybox{8}&1&2&4&6&$c_2$=8\\
&$b=3$&1&3&8&6&2&4&5&7&\mybox{0}&$c_3$=0\\
&$b=4$&6&2&4&5&7&0&\mybox{1}&3&8&$c_4$=1\\
&$b=5$&5&7&0&1&3&8&6&\mybox{2}&4&$c_5$=2\\
&$b=6$&8&1&\mybox{3}&4&6&2&0&5&7&$c_6$=3\\
&$b=7$&\mybox{4}&6&2&0&5&7&8&1&3&$c_7$=4\\
&$b=8$&0&\mybox{5}&7&8&1&3&4&6&2&$c_8$=5\\
\hline
\rule{0pt}{2ex}
&$b=9$&$r_0$=4&$r_1$=5&$r_2$=3&$r_3$=7&$r_4$=8&$r_5$=6&$r_6$=1&$r_7$=2&$r_9$=0&9\\
\end{tabular}
}
\end{table}
The decimal code of Table \wvref{Code47374800} 
is obtained by insertion \wvcite{Damm2007},~\wvcite{Latin1974} 
which projects a transversal \wvcite{Wanless2011} 
to form an added column to the right and an added row at the bottom.  
In Table~\wvref{Code4737} these entries, which will be replaced by `9's,
\iflatexml 
are subscripted with `9's.
\else 
are indicated by boxes.
\fi 
Either $q$, or $\hat{q}$ for emphasis, may denote, here, the additional symbol `9'. 
Transversal entries are not allowed to lie on the main diagonal, and are fixed by
another parameter $P$, e.g. $P=7$ here, with the condition: 
\begin{equation}\label{eq:P}
b\ominus e \: =  \: P \qquad P \in GF(q) \; and \; 0 \neq P 
\end{equation}
Denoting the code entries by $M(b,e)$, the new corner is always filled with $M(q,q)=q$. 
Formulae for the new row $r_e$ and column $c_b$, that also define parameters $R$ and $C$, are:
\mybegineqstar{eq:RC}
\begin{split}
r_e \; = \; K \ominus B \otimes P \oplus e \; = \; R \oplus e \; = \; M(q,e)  \\
c_b \; = \; K \oplus E \otimes P \oplus b \; = \; C \oplus b \; = \; M(b,q)
\end{split}\mytag{\theequation}
\myendeqstar
While $R=C$ is not possible as $R\ominus C=P$, 
that $C\neq 0$ and $R\neq 0$ must be checked to avoid twin errors 
at the start and the end of codewords respectively.
As our field does not have characteristic 2, 
the added row and column are transposition free as permutations because:
\begin{align} \label{eq:NoRorC2cycles}
\begin{split}
M(q,M(q,e)) = R \oplus M(q,e)  = 2 \otimes R \oplus e \neq e  \\
M(M(b,q),q) = C \oplus M(b,q) = 2 \otimes C \oplus b \neq b 
\end{split}
\end{align}
The replacement of an entry in a column of the original $GF(q)$ code 
by the new digit $\hat{q}$ cannot shorten any of the cycles 
when that column is considered as a permutation.  
However, the column could have had an allowed single fixed point 
that would join with the new entry 
in the q\textsuperscript{th} row to form a prohibited cycle of length 2.  
A bit of algebra shows that this  would impliy $C=0$ and similarly $R=0$ for a row.

\iflatexml 
\else 
\begin{savenotes}
\def\notetwo{To run this code in a browser select(drag), on screen and paste into e.g. ASCII \textit{checkDigit.html}.  Optionally, append alphabet size e.g. \textit{?Q=26} to url. 
The code can also be found at the end of the generating LaTeX file at arxiv.org.
Leading or trailing spaces do not affect execution and missing line breaks may be tolerated.
}
\BeginAlgFigure
\caption{Algorithms for 10, 26, and 36 Letter Codes $\color{blue}\uuline{\protect\footnote{\notetwo}}$}\label{alg:html}{\par %
\ttfamily\small
\newcommand\bslash{\symbol{`\\}}
\newcommand\gr{\textasciigrave}
\newcommand\sltr[1]{\textquotesingle#1\textquotesingle}
\vspace{\bufferhline}  
\begin{alltt}
{ <html><head><meta charset="UTF-8">}
{ <style>#OuE\{font-family: monospace;\}</style>}
{ <title>Length 3 Check Digit Codes-ES6</title>}
{ <script>help=()=>hp; Q=10;/*4,9,25,26,36,37*/}
{ var urlstr=document.URL+\gr{}"?Q=$\{Q\}"\gr{};}
{ Q=+urlstr.match(/Q=([0-9]+)/)[1]; Se=\sltr{ };}
{ var _ =undefined, id=x=>x, I=x=>x|0; nl="\bslash{}n";}
{ ta=n=>n<36?(Q%25<2?n+10:n).toString(36):\sltr{-};}
{ AN=0; PK="P"; ts=x=>AN?x:ta(x);Se=\sltr{e};}
{ seq=n=>[...Array(n).keys()]; /*[0,..,n]*/}
{ hp="Add to URL ?Q=N for 4,9,25,26,36,37\bslash{}n";}
{ /*(A)*/ var q=9,G=8,Z=3,K=3,P=7,B=4,E=7;}
{  var pwA = [ 1, 4, 6, 7, 2, 8, 3, 5 ];}
{  var lgA = [-1, 0, 4, 6, 1, 7, 2, 3, 5];}
{ if(Q%25<2)\{G=24;Z=5; K=1;P=1;B=11;E=18; q=25;}
{  mlP=x=>5*((I(x/5)+x%5)%5)+(3*I(x/5))%5; }
{  pwA=seq(G).map(F=i=>(i==0)?1:mlP(F(i-1)));}
{  lgA=seq(q).map(x=>pwA.findIndex(v=>v==x));\};}
{ /*(B)*/ hp+="ad(x,y),ml,sb avail Q=4,9,25\bslash{}n";}
{  ad=(u,v,z=Z)=>(u+v)%z+z*((I(u/z)+I(v/z))%z);}
{  ml=(u,v)=>(u*v==0)?0:pwA[(lgA[u]+lgA[v])%G];}
{  sb=(u,v)=>ad(u,ml((Z-1),v));}
{ Mq=(b,e)=>sb(K,ad(ml(B,b),ml(E,e)));}
{ R=()=>sb(K,ml(B,P)); C=()=>ad(K,ml(E,P));}
{ rwcl=(x,F)=>(x==q)?q:ad(x,F());}
{ M=(b,e)=>(b==q)?rwcl(e,R):((e==q) ?}
{  rwcl(b,C) : ((P==sb(b,e))?q:Mq(b,e)) );}
{ /*(C)*/ ml4=x=>[0,2,3,1][x];}
{ M4=(b,e)=>ad(P,ml4(ad(b%4,ml4(e%4),2)),2);}
{ if(Q==36)\{p7=x=>((x%18==7)?x+18:x)%36;} 
{  M36=(b,e)=>(9*M4(b,e)+64*Mq(b%9,e%9))%Q;}
{  M=(b,e)=>p7(M36(p7(b),p7(e))) \};}
{ if(Q==4)\{M=M4;hp+="Fixed in GF(4): K,B,E\bslash{}n";}
{ ml=(x,y)=>[0,1,2,3,3,,1,,,2][x*y];q=4;Z=2\};}
{ /*(D)*/ if(Q==37)\{ Q=36; K=1; B=10; E=26;}
{ M37=(b,e)=>((K-B*b-E*e)%37+37)%37;}
{  M=M37; hp+="GF(37) K4,B,E are Fixed\bslash{}n";\};}
{ if(Q==9||Q==25)M=Mq; if(Q<5||Q>35)K=(P=1);}
{ if(Q>26||Q==4)PK="K4";if(Q==25||Q==9)PK="_";}
{ /*(E)*/ set=(k=K,p=P,b=B,e=E)=>(K=k,P=p,}
{ B=b,E=e, \gr{}[K,$\{PK\},B,E]=[$\{[K,P,B,E]\}]\gr{});}
{ Te=(b,m)=>(S=Se,(sQ.find(e=>M(b,e)==m))??36);}
{ sQ=seq(Q);SQ=sQ.map(ts);tm=s=>[s,...SQ];}
{ tbl=T=>sQ.map(b=>(sQ.map(e=>ts(T(b,e)))));}
{ csv=t=>t.map(e=>e.join(",")).join(nl);}
{ lbl=t=>tm(S).map((x,i)=>[x,...[SQ,...t][i]]);}
{ show=(f=id,T=M,s=\sltr{m})=>(S=s,csv(f(tbl(T))));}
{ hp+=\gr{}Size $\{Q\}: set() ... set(K,$\{PK\},B,E)\bslash{}n\gr{}}
{ + "m: See Code show() labeled: show(lbl)\bslash{}n"}
{ + "e: show(_,Te)/show(lbl,Te) AN=1 -> #s\bslash{}n"}
{ + "For Q=4,9,25 e.g. show(lbl,ml,\sltr{*})\bslash{}n"}
{ + "MJS()-More JS in myMJS.js";ul="myMJS.js";} 
{ MJS=()=>(js=document.createElement(\sltr{script}),}
{  js.src=ul, document.body.append(js), ul);}
{ V=()=>document.getElementById("OuE").value=}
{  eval(document.getElementById("InE").value);}
{ document.write(\gr{}Current Alphabet Size $\{Q\}\gr{});}
{ </script></head><body>See 2023 arXiv paper}
{ <h4>Alpha. Size 10,26,36 Code Calculator</h4>}
{ <br><input type="text"id="InE"size=99}
{  placeholder="Enter: help(), press = >"/><br>}
{ <input type="button"value="="onclick="V()"/>}
{ <br><textarea id="OuE" readonly}
{  rows=37, cols=99/></textarea></body></html>}
\end{alltt}
\par
} 
\EndAlgFigure
\end{savenotes}
\fi 

The actual entries in the table are then given by the function $M(b,e)$ defined as follows:
\mybegineqstar{eq:Mbe}
M(b,e) = 
  \begin{cases}
      b=\hat{q}&r_e \\
      e=\hat{q}&c_b \\
      P=b\ominus e ~ ~ ~ &\hat{q} \\
      b=e=\hat{q} &\hat{q} \\
      else&M_q(b,e)
   \end{cases}\mytag{\theequation}
\myendeqstar
The final functions in \AlgFigure~\wvref{alg:html}\nbdd-(B), corresponding to equations 
\wveqref{eq:M9}\,--\,\wveqref{eq:Mbe}, compute the entries in the table for $M(b,e)$.  
The functions in \AlgFigure~\wvref{alg:html}\nbdd-(E), 
display the table in response to entering $show()$ or for labelled axes $show(lbl)$. 
A version of the table showing the last digit $e$ as entries 
can be viewed by entering $show(\_,Te)$ or $show(lbl,Te)$.  
These tables can normally be selected, copied from the screen, 
and then pasted into a text file 
resulting in a standard comma separated variable (.csv) formatted output.  
Automation to directly write a file can be obtained by importing your own code 
with the call \textit{MJS()}.
Code is at the end of this .tex source or .html version (\textit{not} the .pdf),
or for minimum functionality with many browsers try \newline
\centerline{\textit{out=t=$>$window.open(encodeURI(\texttt{"}data:text/csv,\texttt{"}+t))}}
as the single line default file \textit{myMJS.js} ($ul=\texttt{"}myMJS.js\texttt{"}$)
in the same folder, allowing $out(show(lbl))$ etc.
The default code size is $Q=10$. 
The underlying $GF(3^2)$ code may be viewed by setting $Q=9$.  
The possibilities developed are $Q\in \{4,\, 9,\, 10,\, 25,\, 26,\, 36,\, 37\}$.
To change $Q$  append e.g. ``?Q=9" to the URL and reselect it,  
or if necessary, edit its value at line~4 of \AlgFigure~\wvref{alg:html} .
Mapping to the alphanumeric characters is performed by function $ta(\_)$.
When $Q\ge 25$, setting $AN=1$ in the input box will result in 
unaligned (no labels in .csv)  numeric output to copy if required.
For $Q\in\{4,\,9,\,25\}$ operation tables can be output, 
e.g. $show(lbl,ml,\texttt{'*'})$ to display a  multiplication table.

\myTwoColSectionvspace
\section{Almost Disjoint Coding}\label{almost}
Multiple codes that each have error detecting properties, 
while being pairwise disjoint can be applicable to the assignment of numbers 
to items in a catalog, banks of lockers, telephone extensions and many other situations.  
For example, a furniture catalog might have items divided into categories 
such as tables, chairs, beds, etc.  
Provided that all numbers for each category are taken from a single code, 
the categories serve as an additional check while all items have a unique number.  
Should dialing a telephone extension at an external branch require first entering a prefix, 
choosing all extension numbers for each branch from a single code may reduce error frequency.  
Section \wvref{OtherBases} will extend the range to possible applications, 
such as vehicle identification tag codes, 
that may require other bases such as 26 or 36.

Creating multiple codes as developed in this paper by using differing parameters 
is a viable solution when a 3\nbdd-digit decimal code is needed. 
However, only 8 disjoint codes can be created in this manner. 
Also, all the codes will contain the codeword \texttt{999} 
and hence will not be fully disjoint. 
The long sought after 4\nbdd-digit twin and transposition error detecting code 
would have generated ten completely-disjoint, 3\nbdd-digit codes, 
one retract for each check digit value, perhaps with equal levels of overall error detection, 
but also perhaps with lesser levels. 

The term \textit{almost-disjoint} will be used here to describe an ensemble of codes 
every pair of which intersect in only a single codeword common to all.  
Thus, it will be possible to have 8 sets of 99 codewords for disjoint coding use.   
To form a set of almost-disjoint decimal codes, $\{D_1, D_2, ...D_n\}$
each with parameters, $[B, \; E, \; K, \; P, \; R, \; C]$, 
the requirements are:

\begin{equation*}
 \begin{array}{lllllr}
 \multicolumn{5}{l}{ \textit{for all~} i, j \; \in \; \{ 1 \ldots n \}  \;\; | \;\; i \neq  j } & ~  \\
~ ~ ~ ~ &D_i.B = D_j.B& ~   &D_i.E = D_j.E  \\
~ ~ ~ ~ &D_i.K \neq D_j.K& ~  &D_i.P \neq D_j.P\\
~ ~ ~ ~ &D_i.R \neq D_j.R& ~  &D_i.C \neq D_j.C \\
~ ~ ~ ~ &D_i.R \neq 0 & ~  &D_i.C \neq 0
\end{array}
\end{equation*}
 
To limit the number of cyclic errors to 9, ( not 27, see Section~\wvref{sec:choose} ) 
it is desirable that each code used also satisfy 
\begin{align}\label{eq:cyclic}
K \neq (B\ominus 1)\otimes P \implies \texttt {cyclic-errors = 9}
\end{align}
While the example code of Table~\wvref{Code47374800} does detect all phonetic errors, 
other parameters result in codes 
with $0,1\,\texttt{or}\,2$ left, and $0,1\,\texttt{or}\,2$ right, phonetic errors.
Using the equations~\wveqref{eq:phonetic} 
the left ($P_L$) and right ($P_R$) phonetic errors can be counted,
\mybegineqstar{eq:phonetic}
\begin{split}
P_L = Card \{ \, e \; \mid \; 0 = M(M(1,e),e) \}  \\
P_R = Card \{ \, b \; \mid \; 1 = M(b,M(b,0)) \}, 
\end{split}\mytag{\theequation}
\myendeqstar
or the pre-calculated codes given in Table~\wvref{seq38} can be used.

Two sequences of almost-disjoint codes are shown in Table~\wvref{seq38}.  
All codes selected have 9 cyclic errors.  
Linear programming was sufficient to search for these sequences 
with the objective of maximizing the number of codes  
while minimizing the number of phonetic errors.  
Unfortunately both of the objectives cannot be attained simultaneously.  
The $B=3, \,E=8$ sequence yields a full sequence of 8 codes 
with the minimum possible total of 16 phonetic errors.  
The $B=4, \,E=7$ sequence is the best obtainable greedy sequence 
with 9 phonetic errors, 
and no greedy sequence can be extended beyond 6 codes.  
A sequence of 7 codes with 12 phonetic errors does exist, 
but simply dropping the last code of the sequence of 8 
given in the table is recommended.
The code of \AlgFigure~\wvref{alg:html} can calculate these codes 
by entering 0 to 4 of the desired parameters as $set(K,P,B,E)$, e.g. $set(1,3)$. 

\begin{table}[tb]
\caption{Two Sequences of Almost Disjoint Codes\mydepthbox}\label{seq38}
\centering%
\myTwoColTableLiftvspace
\setlength{\tabcolsep}{0.5em}
\begin{tabular}{rccccccccclcccccc}
\hline
B&            3&3&3&3&3&3&3&3	  &B          &4&4&4&4&4&4	\\
E&            8&8&8&8&8&8&8&8	  &E          &7&7&7&7&7&7	\\
K&            5&1&2&7&3&4&6&8	  &K          &3&1&5&6&4&2	\\
P&            5&1&2&7&3&4&6&8	  &P          &7&3&2&1&4&5	\\
$P_L$&    0&1&1&1&2&0&2&1	  &$P_L$  &0&0&1&1&1&1	\\
$P_R$&    0&1&1&1&0&2&1&2	  &$P_R$ &0&1&0&1&1&2   \\[0.4ex]
\hline
\end{tabular}\myTwoColTableLiftvspace\end{table}

\myTwoColSectionvspace
\section{Code Parameter Choices and Error Patterns}\label{sec:choose}
The $GF(3^2)$ code always has two phonetic errors, 
in the interior of the example of Table~\wvref{Code4737}, 
left between \texttt{173} and \texttt{703}, and right between \texttt{416} and \texttt{460}. 
These are not necessarily inherited by the decimal extension. 
For example, the decimal extension had codeword \texttt{193} not \texttt{173} and, 
indeed with \texttt{496} in the code, this decimal extension is free of phonetic errors.  
The decimal extension process can also add right and/or left phonetic errors 
when \texttt{b19} and \texttt{b90}, 
and/or \texttt{19e} and \texttt{90e} happen to be codewords.

Given a choice of parameters, $[B,E,K,P]$, 
the reversal of the code generated is produced by parameters $[E,B,K,\ominus P]$.  
The reversed code can vary in the number of phonetic errors.  
However, exchanging the digits `0' and `1'. results in a code 
with again the same number of phonetic errors,
but with left and right phonetic errors exchanged.  
Applying the involution $N(x) = 1\ominus x$ to the original code 
exchanges the digits `0' and `1', 
resulting in a code with parameters $[B,E,\ominus K,\ominus P]$.  
Reversing that code yields a code with parameters $[E,B,\ominus K,P]$ 
that is for most purposes equivalent to the original.  
This observation eliminates $(5,6)$ in the search for parameters $(B,E)$, 
reducing the list to two pairs.

The $GF(3^2)$ code shown in the center of Table~\wvref{Code4737} has 9 cyclic errors.  
The condition for cyclic errors to exist 
between codewords \texttt{abc}. \texttt{bca} and \texttt{cab}, 
multiplying matrices over $GF(3^2)$, is:
\mybegineqstar{cyclic}
\begin{bmatrix}
B & 1 & E \\
1 & E & B \\
E & B & 1
\end{bmatrix}
\otimes
\begin{bmatrix}
a \\
b \\
c
\end{bmatrix}
=
\begin{bmatrix}
K \\
K \\
K
\end{bmatrix}\mytag{\theequation}
\myendeqstar
Note that the left-hand matrix has rank 2 as $0=B\oplus 1\oplus E$ 
and the 9 solutions will divide into 3 groups.
In the case of the code of Table \wvref{Code4737} 
these are the 3 cyclic rotations of each of \texttt{012}, \texttt{345} and \texttt{678}. 

The decimal code will inherit the 9 cyclic errors from the $GF(3^2)$ code 
unless the extension introduces a cyclic error containing the digit `9'.  
Considering the possibility of codewords \texttt{b\,9\,e} and \texttt{9\,e\,b} coexisting, 
one obtains: $b\ominus e=P$ and $e=r_b=K\;\ominus\;B\otimes P\;\oplus\; b$ 
yielding the signaling condition $K=(B\ominus 1)\otimes P$ of \wveqref{eq:cyclic}.  
In our field of characteristic 3, his condition can also be used to show $c_e=b$ 
so that the codeword \texttt{e\,b9} would also exist 
and provably gives at least 27 possible cyclic errors. 
All such codes were examined by a \textit{C++} program and had exactly 27 cyclic errors.

\myTwoColSectionvspace
\section{Alternative 3-digit Codes for Decimal Data}
Of course, codes using a single check digit, designed for longer codewords, 
can be shortened to 3\nbdd-digit length. 
In Table~\wvref{ErrorCounts}, 
the first six codes shown are such shortened versions.  
Only the last three listed codes, all shown in this paper, 
are restricted to detection of errors in 3\nbdd-digit words.  
While the last two codes, developed in this article, 
can be used for disjoint coding within the limits delineated in Section \wvref{almost}, 
any code of the first six listed codes can also serve in this capacity 
yielding 10 completely disjoint 3\nbdd-digit codes.  
The loss of error detecting capability in doing so is made clear by the table.  
The first two codes \wvcite{Damm2007}, \wvcite{Verhoeff1969} use quasigroups 
and the dihedral group of order 10 respectively (see also \wvcite{Gumm1985}).  
The EAN/ISBN-13 and Luhn(UPC-A) codes are in wide use 
and use modulus~10 in their calculations.  
Permutations for use with such modulus~10 calculations have been thoroughly investigated. 
The results of recent searches \wvcite{Alper2017}, \wvcite{Faria2017} for the best modulus~10, 3-permutaion  code 
using differing criteria  are followed in the table by 
the code resulting from use 
of the first three of the sequence of permutations found in \wvcite{Verhoeff1969}.
The latter is labelled ``pg. 81" in the table.  

\iflatexml 
\begin{table}[tb]
\caption{Undetected Error Counts for 3-digit Decimal Codes (*=corrected)\mydepthbox} \label{ErrorCounts}
\centering
\setlength\tabcolsep{2pt}
\begin{tabular}{|ccccccccl|} 
\hline
Single&Transp.&Twin&J-Transp.&J-Twin&Triple&Phon.&Cyclic&Reference\\
\hline
\Z0&\Z0&\Z6&\Z2&\Z3&\Z0&\Z0&12&Damm\,\wvcite{Damm2007}\\
\Z0&\Z0&\Z4&\Z2&2*&\Z0&\Z2&\Z9&Dihedral\,\wvcite{Verhoeff1969}\\
\Z0&10&10&45&\Z5&10&\Z0&\Z0&EAN/ISBN-13\\
\Z0&\Z2&\Z6&45&\Z5&\Z3&\Z2&\Z2&Luhn\,(UPC-A)\\
\Z0&\Z2&\Z5&\Z1&\Z3&\Z0&\Z0&\Z5&\wvcite{Faria2017}\\
\Z0&\Z2&\Z4&\Z1&\Z4&\Z3&\Z1&\Z8&\wvcite{Alper2017}\\
\Z0&\Z2&\Z2&\Z2&\Z2&\Z1&\Z0&\Z5&pg.\,81\,\wvcite{Verhoeff1969}\\
\Z0&\Z0&\Z0&\Z0&\Z0&45&\Z0&16&Irregular\,\wvcite{Verhoeff1969}\\
\Z0&\Z0&\Z0&\Z0&\Z0&\Z0&\Z0&\Z9&Code\,4737\\
\Z0&\Z0&\Z0&\Z0&\Z0&\Z0&0-3&\Z9&Codes\,38xx\\
\hline
\end{tabular}
\end{table}
\else 
\newlength{\toptable}
\begin{table}[tb]
\caption{Undetected Error Counts for 3-digit Decimal Codes} \label{ErrorCounts}
\centering
\myTwoColCaptionvspace
{
\settowidth{\toptable}{EAN/ISBN-13} 
\newcommand{\degslant}{90}
\setlength{\tabcolsep}{0.7em}
\begin{tabular}{@{}ll|ccccccccc|c} 
&Error Types
&\;\;\rotatebox{\degslant}{\makebox[\toptable][l]{\,Damm\,\wvcite{Damm2007}}}
&\rotatebox{\degslant}{\makebox[\toptable][l]{\,Dihedral\,\wvcite{Verhoeff1969}}}
&\rotatebox{\degslant}{\makebox[\toptable][l]{\,EAN/ISBN-13}}
&\rotatebox{\degslant}{\makebox[\toptable][l]{\,Luhn\,(UPC-A)}}
&\rotatebox{\degslant}{\makebox[\toptable][l]{\wvcite{Faria2017}, \wvcite{Alper2017}}}
&\rotatebox{\degslant}
{\makebox[\toptable][l]{\;pg.\,81\,\wvcite{Verhoeff1969}}}&
\rotatebox{\degslant}{\makebox[\toptable][l]{\,Irregular\,\wvcite{Verhoeff1969}}}
&\rotatebox{\degslant}{\makebox[\toptable][l]{\,Code\,4737}}
&\rotatebox{\degslant}{\makebox[\toptable][l]{\,Codes\,38xx}}\\
\hline
\rule{0pt}{3ex}   
&Single&\;\;\Z0&\Z0&\Z0&\Z0&0\,/\,0&0&\Z0&\Z0&\Z0\\
&Transpose&\;\;\Z0&\Z0&10&\Z2&2\,/\,2&2&\Z0&\Z0&\Z0\\ 
&Twin&\;\;\Z6&\Z4&10&\Z6&5\,/\,4&2&\Z0&\Z0&\Z0\\
&J. Transpose&\;\;\Z2&\Z2&45&45&1\,/\,1&2&\Z0&\Z0&\Z0\\
&J. Twin&\;\;\Z3&2*&\Z5&\Z5&3\,/\,4&2&\Z0&\Z0&\Z0\\
&Triple&\;\;\Z0&\Z0&10&\Z3&0\,/\,3&1&45&\Z0&\Z0\\
&Phonetic&\;\;\Z0&\Z2&\Z0&\Z2&0\,/\,1&0&\Z0&\Z0&\Z0-3\\
&Cyclic&\;\;12&\Z9&\Z0&\Z2&5\,/\,8&5&16&\Z9&\Z9\\
\hline
\end{tabular}}\end{table}
\fi 
\myTwoColSectionvspace
\section{Length 3 Codes for Other Bases and Tags}\label{OtherBases}
In this section, a single and permutation error detecting code 
will be said to detect \textit{All} error types.
That will imply, of course, all the types but phonetic, 
which will be noted specifically as appropriate.
In some cases, the almost disjoint construction of Section~\wvref{almost} can be used, 
or completely disjoint coding may be available, 
especially when the base allows a finite field.

Codes derived for smaller bases are of interest as they can be used in e.g. direct product
constructions to produce codes for larger bases.
Computer search using $Prover9/Mace4$ over all possible latin squares has shown that 
the codes for bases $2$--$6$ summarized in Table~\wvref{SmallBases} cannot be improved.
The base 4 code shown in Table~\wvref{Base4} is linear,  
$\alpha \otimes b \oplus m \oplus \beta \otimes e = K_4\neq0$,
or rearranging $M_4(b,e)=K_4\oplus\alpha \otimes(b\oplus \alpha\otimes e)$ over $GF(4)$.
It completely detects all of the error types studied. 
To generate these codes set $Q=4$ in \AlgFigure~\wvref{alg:html} 
and the desired $K_4$ in $set(\_,K_4)$.
\myTwoColTableLiftvspace[-6.0mm]
\begin{table}[h]
\caption{A $GF(2^2)$ Code, $K_4=1$, 
Roots $2=\alpha$ and $3=\beta$ of $x^2+x+1=0$\mydepthbox} \label{Base4}
\centering
\myTwoColCaptionvspace
{
\setlength{\tabcolsep}{0.5em}
\begin{tabular}{l|llll}
M &$\;0$ &$1$ &$2_{\alpha}$ &$3_{\beta}$ \\
\hline
$0$ &$\;1$ &$2_{\alpha}$ &$0$ &$3_{\beta}$ \\
$1$ &$\;3_{\beta}$ &$0$ &$2_{\alpha}$ &$1$ \\
$2_{\alpha}$ &$\;2_{\alpha}$ &$1$ &$3_{\beta}$ &$0$ \\
$3_{\beta}$ &$\;0$ &$3_{\beta}$ &$1$ &$2_{\alpha}$ 
\end{tabular}
}
\end{table}
\myTwoColTableLiftvspace[-0.1mm]
The constructions used to create the $GF(3^2)$ and $GF(4)$ codes, 
e.g. Table~\wvref{Code4737}'s interior code,
can be extended by applying the similar arguments in finite fields $GF(q)$.

\begin{table}[b]
\myTwoColTableLiftvspace
\caption{Small Alphabet Codes\mydepthbox} \label{SmallBases}
\setlength{\tabcolsep}{\mytabcolsepA}
\centering
\myTwoColCaptionvspace
{  
\begin{tabular}{cll}
Base & Source & Errors Detected\\
\hline\\[-2.1ex]
2 & $b\oplus m \oplus e = 0 \mod 2$ & single,~triple\\
2 & $m = 1 \mod 2$ & transpose,~twin\\
3 &$ b\oplus m \oplus e = 0 \mod 3$ & single,~twin\\
3 & $b\oplus 2 \otimes m \oplus e = 0 \mod 3$ & single,~twin\\
4 & Table~\wvref{Base4} \& Theorem~\wvref{divideby3} & All~incl.~permutation\\
5 & $2 \otimes b \oplus m \oplus 2 \otimes e = 1 \mod 5$ & All~but~10~J.~transp\\
5 & $2 \otimes b \oplus m \oplus 3 \otimes e = 1 \mod 5$ & All~but~10~J.~twin,~4 cyclic\\
6 & Table~\wvref{Base6} \& computer search & All~but~5~J.~transp,~5~cyclic\\
$7^n$ & $2\otimes b \oplus m \oplus 4\otimes e = 1,\;  GF(7^n)$& All~incl.~permutation 
\end{tabular}}
\end{table}

\iflatexml 
The codes listed in Table~\wvref{SmallBases} are easily constructed and verified
with the exception of the base~$6$ code of Table~\wvref{Base6} found by computer search.
Beyond base~7 theory currently becomes preferable to complete computer search.
\else 
\fi 
 
\begin{table}[b]
\myTwoColTableLiftvspace
\caption{Base 6: Single, Twin, J. Twin, Transp. and Triple Detecting\mydepthbox}\label{Base6}
\centering
\myTwoColCaptionvspace
{
\setlength{\tabcolsep}{0.5em}
\begin{tabular}{l|cccccc}
M &0 &1 &2 &3 &4 &5 \\
\hline
0 &0 &4 &5 &1 &2 &3 \\
1 &5 &2 &0 &3 &1 &4 \\
2 &1 &5 &3 &0 &4 &2 \\
3 &2 &3 &1 &4 &0 &5 \\
4 &3 &1 &4 &2 &5 &0 \\
5 &4 &0 &2 &5 &3 &1 
\end{tabular}
}
\end{table}

\begin{\myTheorem}[Codes over $GF(q), q=4 \vee q\geq 7$\label{largeq}]
The code over GF(q) with codewords $(b, m, e)$ (denoted $M_q(b,e)=m$)
satisfying $B \otimes b \oplus m \oplus E \otimes e = K$ 
detects all error types except possibly cyclic errors, 
when the conditions of the equations~{\wveqref{eq:BEK}} are met. 
The code will also be triple word free.
\end{\myTheorem}

\begin{\myTheorem}[Permutation Errors\label{permutation}]
A 3\nbdd-digit single error detecting check digit code detects permutation errors if and only if
it detects cyclic and left, right and jump transposition errors.
\end{\myTheorem}

The cyclic errors can be avoided if $3$ divides $q-1$ as follows: 

\begin{\myTheorem}[Permutation Error Free Codes over $GF(q)$\label{divideby3}]
Given a finite field $GF(q)$ where $3 \mid q-1$,
let $\alpha$ be a field element such that $\alpha^3=1$.
Setting $B=\alpha$ and $E=\alpha^2$ in Theorem~\wvref{largeq} 
yields a code satisfying the same conditions,
but with no cyclic errors possible.
\end{\myTheorem}
\begin{\myproof} The multiplicative group of $GF(q)$ is cyclic 
and $\alpha$ generates the subgroup $\{ B, E, 1\}$.
Thus, $B^2\;{=}\;E,\;E^2\;{=}\;B,\;B \otimes E=1$. As $B\neq 1$,  $B\oplus 1 \oplus E=0$ 
follows from:
\begin{align*}(B \ominus 1) \otimes (B \oplus 1 \oplus E) = 0 \end{align*}
The condition for cyclic errors was given in equation~(\wvref{cyclic}). 
It is now easily seen that the columns of the square matrix are all multiples of each other.
As $K\neq 0$, no multiple of these columns equals the right hand side 
to produce a cyclic error.
\end{\myproof}

The simplest example produced by the construction of Theorem~\wvref{divideby3} 
is the $GF(2^2)$ code of Table~\wvref{Base4}.
Note that for $GF(2^3)$ Theorem~\wvref{largeq} must be used 
as $GF(p^d)$ is a subfield of $GF(p^n)$ exactly when $d \mid n$.
Conversely for base~5, note that $3 \mid (24=5^2-1)$ 
and hence Theorem~\wvref{divideby3} can be applied in $GF(5^2)$.
Letting $1=(0, 1)$ and letting $\alpha=(1,0)$ be a root 
of the primitive polynomial  $x^2 + 4x + 2 = 0$ 
the values required by the theorem are $B=\alpha^8=(2,1)=11$, $E=\alpha^{16}=(3,3)=18$, 
and any choice of $K\in GF(5^2), K\neq 0$.
Identifying the entries of $GF(5^2)$ lexicographically 
with the alphabetic characters $a, b, \ldots ,y$ and using $K = 1 = (0,1)$,
gives the code shown in the inner square of Table~\wvref{Base26} 
by viewing the subscripts where applicable.

\begin{table}[t]
\caption{A $GF(5^2)$ Code Extended to Base 26\mydepthbox} \label{Base26}
\centering
\myTwoColCaptionvspace
{
\setlength{\tabcolsep}{0.0em}
\begin{tabular}{c|*{25}{c}|c}
M&a&b&c&d&e&f&g&h&i&j&k&l&m&n&o&p&q&r&s&t&u&v&w&x&y&z\\
\hline
a\;&$b$&$n$&$u$&$h$&$z_t$&$w$&$j$&$q$&$d$&$k$&$s$&$a$&$m$&$y$&$g$&$o$&$v$&$i$&$p$&$c$&$f$&$r$&$e$&$l$&$x$&$t$ \\
b\;&$z_p$&$c$&$o$&$v$&$i$&$l$&$x$&$f$&$r$&$e$&$h$&$t$&$b$&$n$&$u$&$d$&$k$&$w$&$j$&$q$&$y$&$g$&$s$&$a$&$m$&$p$ \\
c\;&$j$&$z_q$&$d$&$k$&$w$&$a$&$m$&$y$&$g$&$s$&$v$&$i$&$p$&$c$&$o$&$r$&$e$&$l$&$x$&$f$&$n$&$u$&$h$&$t$&$b$&$q$ \\
d\;&$x$&$f$&$z_r$&$e$&$l$&$t$&$b$&$n$&$u$&$h$&$k$&$w$&$j$&$q$&$d$&$g$&$s$&$a$&$m$&$y$&$c$&$o$&$v$&$i$&$p$&$r$ \\
e\;&$m$&$y$&$g$&$z_s$&$a$&$i$&$p$&$c$&$o$&$v$&$e$&$l$&$x$&$f$&$r$&$u$&$h$&$t$&$b$&$n$&$q$&$d$&$k$&$w$&$j$&$s$ \\
f\;&$k$&$w$&$j$&$q$&$d$&$g$&$s$&$a$&$m$&$z_y$&$c$&$o$&$v$&$i$&$p$&$x$&$f$&$r$&$e$&$l$&$t$&$b$&$n$&$u$&$h$&$y$ \\
g\;&$e$&$l$&$x$&$f$&$r$&$z_u$&$h$&$t$&$b$&$n$&$q$&$d$&$k$&$w$&$j$&$m$&$y$&$g$&$s$&$a$&$i$&$p$&$c$&$o$&$v$&$u$ \\
h\;&$s$&$a$&$m$&$y$&$g$&$o$&$z_v$&$i$&$p$&$c$&$f$&$r$&$e$&$l$&$x$&$b$&$n$&$u$&$h$&$t$&$w$&$j$&$q$&$d$&$k$&$v$ \\
i\;&$h$&$t$&$b$&$n$&$u$&$d$&$k$&$z_w$&$j$&$q$&$y$&$g$&$s$&$a$&$m$&$p$&$c$&$o$&$v$&$i$&$l$&$x$&$f$&$r$&$e$&$w$ \\
j\;&$v$&$i$&$p$&$c$&$o$&$r$&$e$&$l$&$z_x$&$f$&$n$&$u$&$h$&$t$&$b$&$j$&$q$&$d$&$k$&$w$&$a$&$m$&$y$&$g$&$s$&$x$ \\
k\;&$y$&$g$&$s$&$a$&$m$&$p$&$c$&$o$&$v$&$i$&$l$&$x$&$f$&$r$&$z_e$&$h$&$t$&$b$&$n$&$u$&$d$&$k$&$w$&$j$&$q$&$e$ \\
l\;&$n$&$u$&$h$&$t$&$b$&$j$&$q$&$d$&$k$&$w$&$z_a$&$m$&$y$&$g$&$s$&$v$&$i$&$p$&$c$&$o$&$r$&$e$&$l$&$x$&$f$&$a$ \\
m\;&$c$&$o$&$v$&$i$&$p$&$x$&$f$&$r$&$e$&$l$&$t$&$z_b$&$n$&$u$&$h$&$k$&$w$&$j$&$q$&$d$&$g$&$s$&$a$&$m$&$y$&$b$ \\
n\;&$q$&$d$&$k$&$w$&$j$&$m$&$y$&$g$&$s$&$a$&$i$&$p$&$z_c$&$o$&$v$&$e$&$l$&$x$&$f$&$r$&$u$&$h$&$t$&$b$&$n$&$c$ \\
o\;&$f$&$r$&$e$&$l$&$x$&$b$&$n$&$u$&$h$&$t$&$w$&$j$&$q$&$z_d$&$k$&$s$&$a$&$m$&$y$&$g$&$o$&$v$&$i$&$p$&$c$&$d$ \\
p\;&$i$&$p$&$c$&$o$&$v$&$e$&$l$&$x$&$f$&$r$&$u$&$h$&$t$&$b$&$n$&$q$&$d$&$k$&$w$&$z_j$&$m$&$y$&$g$&$s$&$a$&$j$ \\
q\;&$w$&$j$&$q$&$d$&$k$&$s$&$a$&$m$&$y$&$g$&$o$&$v$&$i$&$p$&$c$&$z_f$&$r$&$e$&$l$&$x$&$b$&$n$&$u$&$h$&$t$&$f$ \\
r\;&$l$&$x$&$f$&$r$&$e$&$h$&$t$&$b$&$n$&$u$&$d$&$k$&$w$&$j$&$q$&$y$&$z_g$&$s$&$a$&$m$&$p$&$c$&$o$&$v$&$i$&$g$ \\
s\;&$a$&$m$&$y$&$g$&$s$&$v$&$i$&$p$&$c$&$o$&$r$&$e$&$l$&$x$&$f$&$n$&$u$&$z_h$&$t$&$b$&$j$&$q$&$d$&$k$&$w$&$h$ \\
t\;&$t$&$b$&$n$&$u$&$h$&$k$&$w$&$j$&$q$&$d$&$g$&$s$&$a$&$m$&$y$&$c$&$o$&$v$&$z_i$&$p$&$x$&$f$&$r$&$e$&$l$&$i$ \\
u\;&$r$&$e$&$l$&$x$&$f$&$n$&$u$&$h$&$t$&$b$&$j$&$q$&$d$&$k$&$w$&$a$&$m$&$y$&$g$&$s$&$v$&$i$&$p$&$c$&$z_o$&$o$ \\
v\;&$g$&$s$&$a$&$m$&$y$&$c$&$o$&$v$&$i$&$p$&$x$&$f$&$r$&$e$&$l$&$t$&$b$&$n$&$u$&$h$&$z_k$&$w$&$j$&$q$&$d$&$k$ \\
w\;&$u$&$h$&$t$&$b$&$n$&$q$&$d$&$k$&$w$&$j$&$m$&$y$&$g$&$s$&$a$&$i$&$p$&$c$&$o$&$v$&$e$&$z_l$&$x$&$f$&$r$&$l$ \\
x\;&$o$&$v$&$i$&$p$&$c$&$f$&$r$&$e$&$l$&$x$&$b$&$n$&$u$&$h$&$t$&$w$&$j$&$q$&$d$&$k$&$s$&$a$&$z_m$&$y$&$g$&$m$ \\
y\;&$d$&$k$&$w$&$j$&$q$&$y$&$g$&$s$&$a$&$m$&$p$&$c$&$o$&$v$&$i$&$l$&$x$&$f$&$r$&$e$&$h$&$t$&$b$&$z_n$&$u$&$n$ \\
\hline
z\;&$p$&$q$&$r$&$s$&$t$&$u$&$v$&$w$&$x$&$y$&$a$&$b$&$c$&$d$&$e$&$f$&$g$&$h$&$i$&$j$&$k$&$l$&$m$&$n$&$o$&$z$
\end{tabular}
}
\end{table}
\par
The insertion method of Section~\wvref{construction} as demonstrated in Table~\wvref{Code4737} 
can be applied to the $GF(5^2)$ code 
to create a base 26 code which will detect all the error types.
The new digit will be denoted by $z$ 
resulting in the single triple word, $z z z$.
With the choice $P=1$, the resulting code appears in Table~\wvref{Base26}.
\begin{\myTheorem} [Code Extension by Insertion\label{insertion}]
Given a code  $M_q(b,e)$ over a field $GF(q)$, not of characteristic $2$,
resulting from Theorem~\wvref{largeq}, 
and given a symbol $\hat{q}$, and $P~\in~GF(q)$, $P~\neq~0$, 
an extended code $M(b,e)$ over $GF(q)\cup \hat{q}$ 
is for $b, e \in GF(q) \cup \hat{q}$ 
is defined by equations (\wvref{eq:RC}) and (\wvref{eq:Mbe}).
The enlarged code $M(b,e)$ will continue to detect all the error types, 
except possibly cyclic as did the code $M_q$, 
but will have the single added triple word $\hat{q} \,\hat{q} \,\hat{q}$.
\end{\myTheorem}
\begin{\myproof}
Suppose code $M_q$ detects all the error types save cyclic, 
then the basic conditions for an enlarging  transversal, $(b_i,e_i), i=1,\cdots q$
are that it be asymmetric 
and that $M(b_i,e_i) \notin \{b_i, e_i\}$, i.e. $R\neq 0, \; C\neq 0$. 
With these conditions, the only remaining possible errors 
would be due to transpositions in the added row or the added column.
Suppose there is a transposition error $x\, y\, \hat{q} \leftrightarrow  y\, x\, \hat{q}$.
Let $x\, y\, u$ and $y\, x\, v$ be the original codewords that were overlaid with $\hat{q}$.
It follows that $x\ominus u = P = y \ominus v$, 
and using $B \oplus 1 \oplus E = 0$ to manipulate 
\begin{align*} 
B \otimes x \oplus y \oplus E \otimes u = K = B \otimes y \oplus x \oplus  E\otimes v 
\end{align*}
results in concluding $2\otimes (x \ominus y)=0$ and hence $x=y$. 
The case for the added row is similar.
\end{\myproof}
\begin{\myTheorem}[Permutation Error Free Codes by Insertion\label{cyclicInsert}]
When a code is produced by Theorem~\wvref{divideby3} and is not of characteristic $2$,
the insertion method of Theorem~\wvref{insertion} 
yields an extended code that also detects all cyclic errors provided that 
\begin{align*}
K\neq (B\ominus 1)\otimes P\;  \land \; K \neq (1 \ominus E)\otimes P
\end{align*}
\end{\myTheorem}
\begin{\myproof}
Any added cyclic error must involve the added symbol $\hat{q}$ 
appearing in both the middle and an end position.
Consider the case where $b\, \hat{q}\, e$ and $e\, b\, \hat{q}$ are codewords. 
Then $b \ominus e = P$ 
and $b = K \oplus E\otimes P \oplus e = K \oplus E\otimes P \oplus b \ominus P$ 
and hence $ K = (1 \ominus E)\otimes P$.  
For the case where $ b\, \hat{q} \,e$ and $\hat{q}\, e\, b$ are codewords 
$K = (B \ominus 1)\otimes P$ as developed in Section~\wvref{sec:choose}.
\end{\myproof}
\begin{\myTheorem}[Codes for the 26 Letter Alphabet\label{code26}]
A base 26 code that detects all error types, and has a single triple word can be produced
by inserting an additional symbol `$z$', setting $B=(2,1)$, $E=(3,3)$, 
and choosing parameters $K=P \in GF(5^2)\neq 0$. 
This results in 24 almost disjoint codes having only the codeword $z\, z \, z$ in common.
\end{\myTheorem}
\begin{\myproof} The parameters for each code meet the conditions of Theorem~\wvref{cyclicInsert}
noting that $R = K\ominus B\otimes P=(1\ominus B)\otimes K \neq 0$ as $B\neq 1$
and similarly for $C\neq 0$. 
To show two codes to be almost disjoint, note that the differing values of $K$ 
place them in different cosets except for the values changed by insertion.
The differing values of $P$ guarantee that the new symbol `$z$' 
is overlaid only to form the codeword $z \, z\, z$.
The parameter $R$ differs
as $K=P$, $R=(1\ominus B)\otimes K$ and $1 \ominus B \neq 1$, and similarly for $C$.
\end{\myproof}

\AlgFigure~\wvref{alg:html} can generate these alphabetic of Theorem~\wvref{code26} codes 
with insertion by setting $Q=26$ or the $GF(5^2)$ codes with $Q=25$.
For information on \textit{sum composition} methods adding 
more than one extra row and column see \wvcite{Hedayat78}, \wvcite{Ruiz74}.

The preceding theorem can be applied to generate base~$26$ codes with 
codewords having two (or more) groups of three letters.
Denoting  the codes from Theorem~\wvref{code26} by $C_{K,P}$, 
and identifying $\{0, 1, \cdots 23\}$ with the nonzero members of $GF(5^2)$,
two identical sets of 3\nbdd-digit codes $\mathcal{C}^0, \; \mathcal{C}^1$ are formed:
\begin{align*}
\{\mathcal{C}^i_0, \cdots \mathcal{C}^i_{23}\}  = \{C_{K,K}\setminus(zzz) \mid 0\neq K \in GF(5^2)\}.
\end{align*}
Now, making use of the single-error detecting code 
$\mathcal{S} = \{(0,1), (1,2), \cdots, (22,23), (23,0) \}$ set:
\begin{align*} 
\mathcal{C}= \{(c_0, c_1) \mid c_0 \in \mathcal{C}^0_i,\; c_1 \in \mathcal{C}^1_j,\; (i,j) \in S \} 
\end{align*}
The code $\mathcal{C}$ has $(26^2-1)^2\times 24$ 6\nbdd-digit codewords.
Each 3\nbdd-digit group has the protection provided by Theorem~\wvref{code26} 
given that there is error in only one group.
The code $\mathcal{S}$ is transposition error detecting, rendering the code $\mathcal{C}$
group reversal error detecting. 
The code provides $92\%$ of the $26^5$ codewords possible using a single check symbol.
Allowing retention of the triple word $z\,z\,z$ in the two component codes $\mathcal{C}^i_0$
would yield a slight increase to $10 \; 936 \; 351$ codewords.
Using base~$10$ and $k=8$ results in only $78\%$ of possible. Generalizing:

\begin{\myTheorem}[The Tag Code  or License Plate Theorem\label{licenseplate}]
Let $\mathcal{C}^0,\cdots,\mathcal{C}^{\ell-1}$ be collections each of $k$ disjoint codes 
of lengths $n_0,\ldots,n_{\ell-1}$
over alphabets $\mathcal{A}_0,\ldots,\mathcal{A}_{\ell-1}$ respectively.
Let the error patterns detected by every code $C \in \mathcal{C}^i$ 
be denoted by $\mathcal{E}_i$.
Denote the codes in these collections by $\mathcal{C}^i = \{C^i_0, \cdots, C^i_{k-1}\}$.
Let $\mathcal{S} \in \mathbb{Z}_k^{\ell-1}$ be a single error detecting code.
The concatenation $\mathcal{C}$ of codewords:
\begin{align*}
\mathcal{C} = 
\cup_{i=0}^{k-1} \{ (c_0,\ldots,c_{\ell-1} ) \mid s \in S,\, c_i \in C^i_{s_j} \} 
\end{align*}
detects the error types $\mathcal{E}_i$ in the $i^{th}$ component codeword 
provided that only one component is in error.
\end{\myTheorem}

\begin{table}[htb]
\caption{Grouped 6\nbdd-digit Codes from Theorem \wvref{licenseplate}\mydepthbox} \label{grouped}
\centering
\myTwoColCaptionvspace
{
\setlength{\tabcolsep}{\mytabcolsepA}
\begin{tabular}{@{}l|*{4}{c|}}
&Alphabet 
&Total Words 
& Sample Word 
&Errors Covered \\
& & \% of Bound & \# Codes & Notes 
\\[\mytabcolsepA]
\hline
\rule{0pt}{2.2ex}   
&10& 78\,408 &4\,0\,5\; 9\,0\,8 &  All but Cyclic, Phon \\
&0,\ldots, 9& 78\% & $k=8$ & 9\,9\,9 not used \\[\myrowskip]
\hline 
\rule{0pt}{2.2ex} 
&26& $10\,936\,351$ & z\,w\,h\; j\,x\,z & All incl. Perm \\
&a,\ldots, z& 92\% &  $k=24$ &  \\[\myrowskip]
\hline
\rule{0pt}{2.2ex} 
&36& $58\,786\,560$ & a\,8\,3\; 9\,z\,i & All incl.  Perm + Phon \\
& 0,\ldots,9,a,\ldots , z& 94.6\% & $k=35$  &  mod $37$ code \\[\myrowskip]
\hline
\rule{0pt}{2.2ex} 
&36 & 40\,310\,784 & 1\,1\,x\; 1\,x\,0 & All incl. Perm + Phon \\
 & 0,\ldots,9,a,\ldots , z  & 75\% & $k=27$ & $GF(9): B=4, E=7$  \\[\myrowskip]
\hline
\end{tabular}
}
\end{table}

A direct product construction can be used to combine a 
code~$C_1$ of base~$Q_1$ with a code~$C_2$ of base~$Q_2$ 
resulting in a product code $C = C_1 \myOtimes C_2$ of base~$Q_1\times Q_2$.  
For each pair of codewords 
$(b_1\, m_1\, e_1)=c_1\in C_1$ and $(b_2\, m_2\, e_2)=c_2 \in C_2$,
$C$ will then contain the word $(b_1,b_2)\, (m_1,m_2)\, (e_1, e_2)$.
\begin{\myTheorem}[Product Construction\label{product}]
Given 3\nbdd-digit codes $C_1$ and $C_2$ suppose that:
\begin{enumerate}
\item[] $C_1$ detects all error types and has no triple words.
\item[] $C_2$ detects all error types except possibly triple and cyclic errors.
\end{enumerate}
Then the 3\nbdd-digit product code $C_1 \myOtimes C_2$ has no triple words and detects 
all the error types including all permutation errors.
\end{\myTheorem}
\begin{\myproof}
A triple word in the product code must be the result of triple words in both codes.
A cyclic error in the product code must be due to cyclic errors in both codes
or to a cyclic error in one code and a triple word in the other.
While a transposition or twin error can be the result of two positions changing in one code
and being identical in the other, 
it must commence with such an error in one code or the other.
\end{\myproof}

Alphanumeric applications naturally lead to consideration of base $36$ codes.
Let $C_1$ be the $GF(4)$ code shown in Table~\wvref{Base4},
and let $C_2$ be the $GF(9)$ code shown in the interior of Table~\wvref{Code4737}.
The resultant product code is a base~$36$ code 
that suffices to contain the 26-letter alphabet and the decimal digits,
while detecting all error types. 
The \textit{Chinese Remainder Theorem} allows 
$f(b,e)=(9\,(b\, mod\, 4)+64\,(e\, mod\, 9))\, mod\, 36$ 
as a mapping since 4 and 9 are relatively prime.
By varying the values of $K$ and $K_4$, $set(K,K_4)$, with alphabet size $Q=36$ 
and using \AlgFigure~\wvref{alg:html}\nbdd-(C), this generalizes as follows:

\begin{\myTheorem}[Disjoint Codes for a 36 Letter Alphabet\label{Base36}]
Let $C_4^i$ be the code constructed over $GF(4)$
using $B_4=\alpha, \; E_4=\beta, \; K_4=i\neq 0$,
and let $C_9^j$ be the code constructed over $GF(9)$
using $B_9=4, \; E_9=7,\; K_9=j$.
Then
\begin{align*}C_{i,j}  \overset{\Delta}{=} C_4^i \myOtimes C_9^j \end{align*}
is a 3\nbdd-digit code detecting all error types 
for any of the $3\times 9 = 27$ fixed choices of $i\neq0,\,j$.
Moreover, these codes are disjoint from each other 
and detect all phonetic errors after exchanging
the characters $7$ and $p\leftrightarrow 25$ (choice used in Algorithn~\wvref{alg:html}-(C)).
\end{\myTheorem}
\begin{\myproof} 
For fixed choices of $i\neq 0 \in GF(4), \,j \in GF(9)$ 
the conditions of Theorem~\wvref{product} are met.
Setting $j=0$  alows triple errors in $C_9^0$ only.
As the codes $C_4^i$ are disjoint cosets of each other, 
any two product codes with differing values of $i$ must be disjoint,
and similarly for~$j$.
As phonetic errors are independent of $K_4$ and $K$, generating any code table shows that 
all ``phonetic errors" with $B = 4$ and $E = 7$ have the patterns
$\_\, 1\, f \leftrightarrow  \_\, f\, 0$ where $f \leftrightarrow 15$, 
and $1\, 7\, \_ \leftrightarrow   7 \, 0\, \_\,$.
\end{\myproof}

Theorem~\wvref{divideby3} can be applied to generate base~$37$ codes
using ordinary modulus $37$ calculations with $B=10$ and $E=26$ and $K\in\{1,\ldots ,36\}$.
\AlgFigure~\wvref{alg:html}\nbdd-(D) defaults to $K=1$; use $\,set(K)\,$ to change.
The elimination of all codewords containing the symbol corresponding to $36$ 
hampers inclusion of meaningful key data, but
will produce a code with $97.3\%$ of the $36^2$ possible codewords.
Subsequently using Theorem~\wvref{licenseplate} would then result in a code
with $94.6\%=97.3\%^2$ of the upper bound  $36^5=60\,466\,176$ codewords.
These codes detect all phonetic errors 
as exchanging $1X$ and $X$0 implies $X=34\leftrightarrow \texttt{y}$ 
right or left and for any choice of $K$. 
\AlgFigure~\wvref{alg:html}\nbdd-(D) can generate these codes by setting $Q=37$.
\iflatexml 
The alphabet size will, of course, remain $36$.
\else 
\fi 

For longer lengths, 
the coefficients for extending alphanumeric codes with these fields are easily generated:
\begingroup
\addtolength{\jot}{-0.5mm}
\begin{align} 
&GF(4): {a_1=\alpha, a_{i+1}=a_i\otimes \beta} \nonumber \\ 
&GF(9): {a_1=4, a_{i+1}=a_i\oplus 6}  \nonumber \\ 
&GF(37): {a_1=10,\;  a_{i+1}=a_i \times 26 \; (\;mod\; 37)} \nonumber
\end{align}
\endgroup

\enlargethispage{\equalizelastpgcols}

\myTwoColSectionvspace
\section{Remarks}

An objective of this paper has been 
to allow the codes discovered to actually be used with reasonable effort.
\iflatexml 
Given the tables provided, and the table generating code, in many cases, 
no programming should be required as the uses are generally static 
with one-time assignment of codewords. 
\else 
Given the tables provided$\color{blue}\uuline{\protect\footnotemark}$, and the table generating code, in many cases, 
no programming should be required as the uses are generally static 
with one-time assignment of codewords. 
\fi 

While this article has been restricted to single check digit schemes, 
codes using more that one check digit 
to provide enhanced error detection, or correction Brown \wvcite{DAHBrown1974}, 
have been studied, including modulus 11~Beckley \wvcite{Beckley1967}, 
and modulus 97 Briggs \wvcite{Briggs1971}, Putter and Wagner \wvcite{Putter1989} schemes. 
Codes detecting insertion, deletion and run length (shift) errors 
normally require multiple positions containing check data, 
which can be mixed with payload data and thus may not be \textit{systematic}, 
Tang and Lum \wvcite{Tang1970}, Dunning and Arora \wvcite{Dunning2002}, Briggs \wvcite{Briggs1970}.

Consideration of the many well-designed codes given in the references is recommended 
when a codes with groupings of 4 digits or more is needed.  
The text by Kirtland \wvcite{kirtland2001} provides an introductory survey.  
Alper \wvcite{Alper2017} compares 
the Damm \wvcite{Damm2007}, Verhoeff  dihedral \wvcite{Verhoeff1969} and Luhn(UPC-A) codes.  
Codes for bases other than 10 
are developed in Gumm \wvcite{Gumm1985}, and Damm \wvcite{Damm2007}.  
Belyavskaya et al. \wvcite{Quasigroup2017}  explores other bases as well, 
while  Abdel-Ghaffer \wvcite{Abdel1998} also gives an extensive bibliography. 

Although check digit codes are of mathematical interest, 
the error types and bases of practical interest have and continue to change.  
The difficulty of bases of the form $Q=4n +2$ has a history 
beginning with codewords for telegraph messages Friedman~\wvcite{Friedman1932} $Q=26$, 
and then $Q=10$ with telephone dialing considertations Rothert \wvcite{Rothert1963}.  
Putter and Wagner \wvcite{Putter1989} 
give advice on procedures for selecting a code for a client.  

\iflatexml 
\section{Endnotes}

\begin{enumerate}
\item\label{foot1}\par{When referring to the original tables on pages 40-41 of \wvcite{Verhoeff1969}, 
the code shown inTable~\wvref{VerhoeffRegular}, ``resulting" from the block design, 
is erroneously interchanged with the ``interchanged" code (not shown) 
in which the codeword ``\texttt{109}" appeared as ``\texttt{100}".}

\item\label{foot2}\par{To run the code form Algorithm~\wvref{alg:html} in a browser select(drag), on screen and paste into e.g. ASCII (UTF-8) file \textit{checkDigit.html}.  
Optionally, append alphabet size e.g. \textit{?Q=26} to url. 
A copy of a more capable output routine myMJS.js to be loaded with call MJS() appears in an appendix.
The Algorithm~\wvref{alg:html} code can also be found in the PDF
or with additional auxilliarly code at the end of the 2024 generating LaTeX file source at arxiv.org.
Leading or trailing whitespace does not affect execution and missing line breaks may be tolerated.

}

\item\label{foot3}\par{\copyright The author grants copyright permissions as specified in
Creative Commons License CC BY-SA 4.0.  
Attributions should reference the PDF version of this manuscript, which is preferred for printing.
This HTML version is typeset with current program code as of \today.
The author may be reached via e-mail:dunning@bgsu.edu at Larry A. Dunning, Professor Emeritus, 
Department of Computer Science, Bowling Green State University, Ohio 43403, USA.}
\end{enumerate}

\else 
\fi 


\iflatexml 
\else 
\vfill
\hrule
\footnotetext{An "*" marks an entry corrected from an earlier version in Table~\wvref{ErrorCounts} . }
\fi 

\iflatexml 
\onecolumn
\appendices
\section{File Output Code}

The code shown below allows output to be written to a file with a given name.
It functions with a greater range of browsers and readers
than the one line solution given in the main text. 
An alternate output routine with automatic file naming 
can be found in comments near the end of the 2024 .tex file, 
but does not appear here or in the .pdf files.

\begin{algorithm}[Code for myMJS.js  to Output Named Files\textsuperscript{\wvref{foot2}}]
\end{algorithm}
{\par %
\ttfamily
\newcommand\bslash{\symbol{`\\}}
\newcommand\gr{\textasciigrave}
\newcommand\tqs{\textquotesingle}
\newcommand\sltr[1]{\textquotesingle#1\textquotesingle}
\begin{alltt}
{/* COMMAND help() AFTER LOADING THIS DYNAMICALLY LOADED, ADD-IN BY ENTERING: MJS() */}
{/* This Javasccript code is written to be imported into a specific web page.       */}
{/* That html/javascript page can be found via the pdf or html manuscript           */}
{/* "Length 3 Check Digit Codes with Grouped Tags and Disjoint Coding Applications" */}
{/* at arxiv.org "https://doi.org/10.48550/arXiv.2312.12116" for details on use.    */}
{/* Import is accomplished by entering the call MJS() from the specific web page.   */}
{/* This file can be given the default name myMJS.js or enter ul="myFilename.js"    */}
{/* Verifcation routines L3CVerify.js are currently available from dunning@bgsu.edu */}
{/* Original publication space restrics the javascript facilities that can be used. */}
{/* */}
{if (typeof oldsaveHp == "undefined") \{ var oldsaveHp = hp;}
{hp=\tqs{}To ouput to a file replace show(...) with fileit("myFilename",...)\bslash{}n\tqs{}+oldsaveHp\};}
{function fileit(filename,fsave=id,Tsave=M,sSave=\sltr{m})\{}
{    csvOutput=show(fsave,Tsave,sSave);}
{    let csvContent = "data:text/csv;charset=utf-8,";}
{    var encodedUri = encodeURI(csvContent+csvOutput);}
{    var downLoader = document.createElement(\sltr{a});}
{    downLoader.setAttribute("href", encodedUri);}
{    downLoader.setAttribute("download", filename + ".csv");}
{    var downButton = document.createElement("button");}
{    downButton.innerHTML="Download " + filename + ".csv";}
{    function RemoveChildren(ChList) \{ChList.map(c=>document.body.removeChild(c));\};}
{    function RemoveOur3() \{RemoveChildren([cancelButton,downButton,downLoader])\};}
{    downButton.onclick = function() \{ downLoader.click(); RemoveOur3(); \};}
{    var cancelButton = document.createElement("button");}
{    cancelButton.innerHTML="Cancel";}
{    cancelButton.onclick = function () \{ RemoveOur3(); \};}
{    document.body.appendChild(downLoader);}
{    document.body.prepend(cancelButton);}
{    document.body.prepend(downButton);}
{    return csvOutput;\};}
{InputEntry = document.getElementById("InE");}
{InputEntry.placeholder = "fileit(...) loaded, Enter: help(), press = >";}
{setTimeout ( () => \{ InputEntry.value=""; \}, 100 );}
{/* */}
\end{alltt}
\par
} 
\noindent\hrule
\else 
\fi 

\iflatexml 
\else 
\fi 

\end{document}